# Dual-Quaternions: Theory and Applications in Sound

*Benjamin Kenwright*

### Abstract
Sound is a fundamental and rich source of information; playing a key role in many areas from humanities and social sciences through to engineering and mathematics. Sound is more than just data 'signals'. It encapsulates physical, sensorial and emotional, as well as social, cultural and environmental factors. Sound contributes to the transformation of our experiences, environments and beliefs. Sound is all around us and everywhere. Hence, it should come as no surprise that sound is a complex multicomponent entity with a vast assortment of characteristics and applications. Of course, an important question is, what is the best way to store and represent sound digitally to capture these characteristics? What model or method is best for manipulating, extracting and filtering sounds? There are a large number of representations and models, however, one approach that has yet to be used with sound is dual-quaternions. While dual-quaternions have established themselves in many fields of science and computing as an efficient mathematical model for providing an unambiguous, un-cumbersome, computationally effective means of representing multi-component data. Sound is one area that has yet to explore and reap the benefits of dual-quaternions (using sound and audio-related dual-quaternion models). This article aims to explore the exciting potential and possibilities dual-quaternions offer when applied and combined with sound-based models (including but not limited to the applications, tools, machine-learning, statistical and computational sound-related algorithms).

### Introduction

Sound data and the associated processes for managing, manipulating, combining, filtering and generating are complex and diverse (see Figure 1). When we talk about 'sound' it is more than just 'noise', it is about making sense of the 'sound' data. Sound is as much noise as an image is just colors. Sound is a key part of the data-revolution explosion; with vast amounts of audio data being created, collected and monitored constantly. In order to address the challenges and limitations of sound-based models of the future, we must explore and develop new models and ways of thinking; one such idea is dual-quaternions [9].

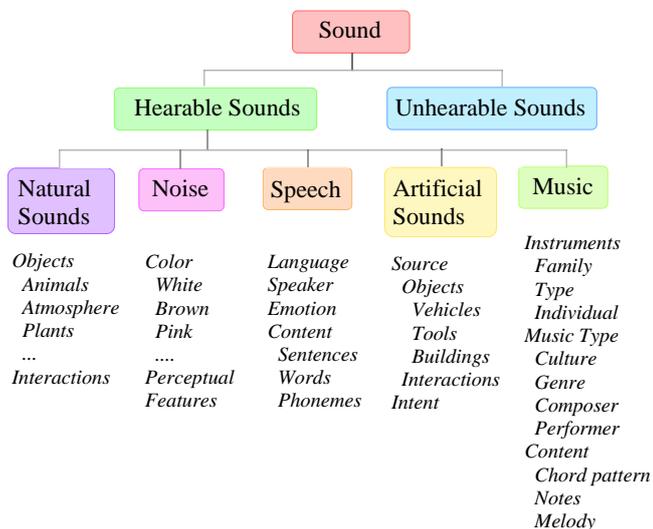

*Figure 1: Overview of the diverse range of sound categories (see [16] for a comprehensive survey of the taxonomy of sound types).*

Dual-quaternions are simply the unification of dual-number theory with hypercomplex numbers. This mathematical concept allows multi-variable data sets to be transformed, combined, manipulated and interpolated in a unified non-linear manner. This non-linearity is critical in many sound models, for example, reducing degeneracies during user editing/compression. Additional transforms could be applied to the various sound/audio frequencies to influence the level of discretization, making it possible to model both coarsely and densely sampled distributions. To highlight some of the novel dual-quaternion concepts proposed in this article:

- Encoding, representing and transforming sound/audio signals to/from dual-quaternion forms
- Impact on sound quality/sound aesthetics
- Analysis and intuition of sound distribution models (dual-quaternion space)
- Non-linear sound space formulations
- Algorithm mapping between sounds signals
- Application of dual-quaternions in interactive editing of sounds/audio (computational speed factors)
- Dual-quaternion/sound formulations in differentiable deep learning architectures

The article aims to underline the benefit from a dual-quaternion in sound. Benefits that go beyond minor incremental improvements but a whole new way of working and seeing sound data - with uses and applications in the wider context - educating and helping us solve problems. Of course, this raises questions, such as how the models and approaches developed in this article will impact other areas of the tool chain. Sound/audio is an important element in other areas beyond 'listening' but other concepts, like sensors (sonar).

**Key Contributions**: The key contributions of this article are: (1) we survey and propose dual-quaternions as a key resource for sound-based models (including but not limited to the applications, tools, statistical and computational model); (2) comparison and review of current and future sound related challenges and trends; (3) a key function of this article is to facilitate how dual-quaternions can add value to sound-based concept and applications through explorative/pilot research.

## Dual-Quaternion Sound Applications

The dual-quaternion algebra is very convenient method for representing audio and related attributes, especially for complex multicomponent signals. The basic dual-quaternion mathematics and theory is detailed and explained in Kenwright [3,13]. We describe several novel applications of dual-quaternions in sound-based algorithms, processes and models. One huge advantage of dual-quaternions is the fact that we can use them for transformations, which is not the case with other formalisms.

The applications of the dual-quaternions are enormous [4] and have the added advantage of being able to leverage related discoveries from other fields that have used dual-quaternions or related research (e.g., from signal filtering) with potential applications such as:

- Encoding/decoding (noise-reduction)
- Filters and pattern identification/creation/removal [2]
- Error correction and compression [3]
- Security (adding hidden tags to audio signals, akin to the work done by Y Chen et al. who developed a quaternion color watermarking technique[1])
- Machine learning [5] extending to non-real number models
- Sound can provide important clues such as sound directionality and spatial size
- Connect with various sound synthesis methods, including harmonic synthesis, texture synthesis, spectral analysis, and physics-based synthesis
- Sound rendering [11] (multi-resolution [10])
- Hybrid methods (mix and match dual-quaternions with other concepts to find a suitable balance and fit)

## Machine Learning

Artificial neural networks (ANNs) based machine learning models and especially deep learning models have been widely applied in signal processing and many other domains, where complex numbers occur either naturally or by design. However, most of the current implementations of ANNs and machine learning frameworks use **real numbers** rather than other representations, such as, dual-quaternions. There are growing interests in building ANNs using other number types, such as Complex Numbers (CVNNs) [5], and exploring the potential advantages over their real-valued counterparts. We discuss the recent innovations and advantages around these works in the literature and their suitability with dual-quaternion configurations. Specifically, a detailed review of dual-quaternion networks in terms of activation functions, learning and optimization, input and output representations, and their applications in audio tasks like signal processing followed by a discussion on some pertinent challenges and future research directions.

## Dual-Quaternionic Signals

Discrete signals can be converted to quaternion space [14,7] known as 'quaternionic signals'. We propose extending this to dual-quaternion-based signal analysis (DQSA) methods (dual-quaternionic signals). Examples include, but are not limited to, dual-quaternionic correlation for vector signals, dual-quaternionic DSVD (DSVDQ) for vector-sensor array, dual-quaternion icwavelet transform (DQWT) for audio signals, dual-quaternionic adaptive filtering, dual-quaternionic independent component analysis, and blind extraction of dual-quaternionic sources.

## Dual-Quaternion Fourier Transform

The Quaternion Fourier transforms (QFT) [12,7] concept is able to be extended to create a Dual-Quaternion Fourier Transform (DQFT) which provides a powerful tool for the analysis of signals with multiple attributes. Signal acoustics enable the dual-quaternion representation to be transformed to a time-frequency domain based. Note, while the standard Fourier analysis and methods are a powerful tool, when they're extended and combined with other mathematical models their benefits are magnified and encapsule new paradigms. We refer the readers to [12,7,13] for further details on these topics.

Sound research has made great progress in recent years [9]. Many methods have been developed that have achieved excellent results in both sound quality and computational efficiency, and some of these methods have even been applied in online content, computer games, virtual reality and films.

Quaternions and Dual-Quaternions have been successful in multiple fields including animation [7], inverse-kinematics [6], geometric representations like curves and surfaces [8].

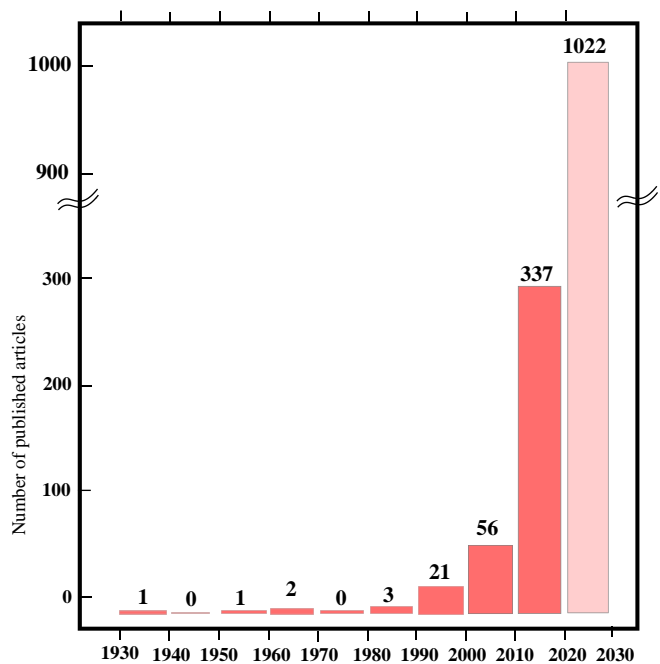

*Figure 2: Visually illustrating the rising popularity and impact of dual-quaternions by showing the number of articles with the word 'dual-quaternion' in the title published between 1930 and 2020 (Google Scholar – Accessed 20/01/2020). Prediction for the years ahead to 2030.*

## Taxonomy

A high-level taxonomy of sound showing the structure of sound related processes/models/algorithms helps visualize the broad range of opportunities dual-quaternion based modifications (see Figure 3). We also list a few opportunities:

- Coupling sounds with positional data (or rotational) data
- Artistic factors – unique audio sounds with bespoke characteristics
- Environmental factors (medium) – as sound is the rapid cycling between compression and rarefaction of 'air' – which is linked with 'rotational' frequencies for specific tones
- Dual-quaternions have been used to represent 'position' and 'rotation' of rigid bodies – however, for 'sound' they could be used to represent other parameters (not just physical components but other underlying information)
- Combining and manipulating dual-quaternion sound wave representations (e.g., superposition)
- Spectral transformations of dual-quaternion representations of the signals (e.g., frequency domain)
- Analytic sound expression for mapping specific characteristics to corresponding generators (e.g., water drops or fluid flow)
- Sound synthesis (model attempts to characterize sound textures)
- Acoustic functions
- Sound noise (generation, type and distribution)
- Interpolation and mixing of sounds [13,8,17]

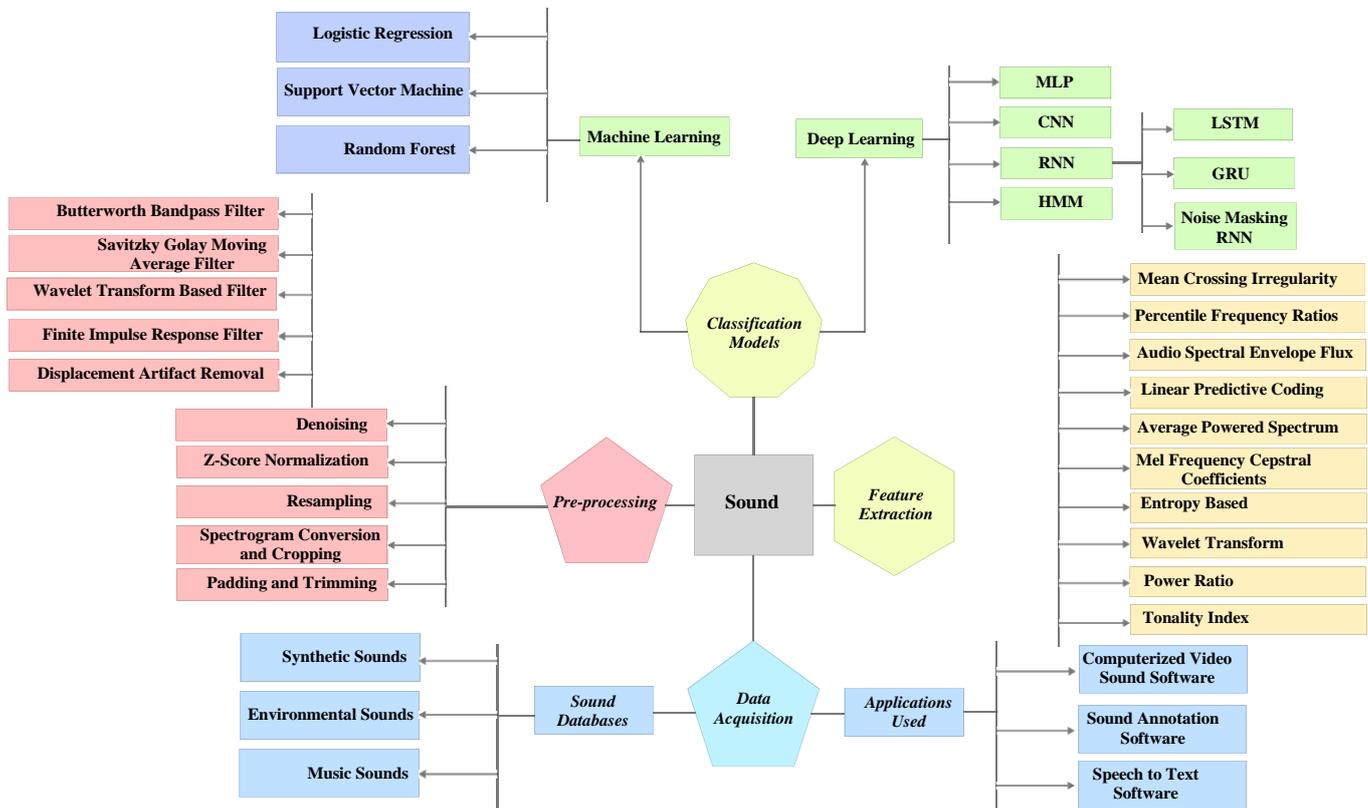

*Figure 3: Brief taxonomy of sound showing the different areas dual-quaternion can have an impact (pre-processing/filtering through to machine learning and classification).*

### Discussion

A main advantage of dual-quaternions is that we can transform and combine several signals and associated attributes. Therefore, it is quite easy to compute and write the expression for complex systems. We used that advantage throughout this article to justify novel applications. Another advantage of using dual-quaternions to represent signals is that it is an efficient way to parameterize characteristics and signals (without singularities and in a non-linear way). There are also factors around memory usage and computational (dual-quaternions offer a efficient and compact representation which could be designed to leverage hardware acceleration).

Dual-quaternions have been widely used to model translational and rotational components [3,8,13]. Here, we described how the dual-quaternion formalism can be used to model the multi-dimensional aspects of sound, which can be useful for modelling purposes in applications like analytics, security and filtering.

This article has explored and justified the value and potential benefits of dual-quaternions in sound models; with applications ranging from improved filtering through to new machine learning paradigms. Dual-quaternions are a clear game-changing idea for sound which would leading to new products, processes and services. To the best of our knowledge, this is the first attempt to connect and apply sound research with dual-quaternions. There are plenty of challenges in terms of far-field automatic speech recognition, source separation, localization in noisy environments.

Dual-quaternions have been commonly used in robotics and computer vision, for various purposes [4,6,13,3]. These concepts developed in solving problems in other areas may be modified and extended to the area of sound and audio models. Dual-quaternions allow sound data to be represented in a form for extracting, filtering, storing and more. They provide and present a new set of rules (and mathematical laws) for managing, controlling and creating sound in new and unimaginable ways.

| Sound Quality | Bandwidth | Sampling rate | Number of bits | Data rate (bits/sec) | Comments |
|---|---|---|---|---|---|
| High fidelity (e.g., music) | 5Hz to 20kHz | 44.0kHz | 16 bit | 700k | Better than human hearing. |
| Telephone quality speech (with companding) | 200Hz to 3.2kHz | 8kHz | 12 bit | 96k | Good quality, but very poor for music |
| | 200Hz to 3.2kHz | 8kHz | 8 bit | 64k | Nonlinear encoding reduces the data rate by 50%. |
| Speech encoded by Linear Predictive Coding | 200Hz to 3.2kHz | 8kHz | 12 bit | 4k | DSP speech compression technique. Very low data rates, poor voice quality. |

*Table 1: Show popular sound quality attributes based on the application (e.g., see Figure 1), such as, bandwidth, number of bits and quality.*

In summary, dual-quaternions are powerful mathematical constructs that are widely used in robotics and computer vision, and could make important contributions to sound research.